\documentclass{iaus}
\usepackage{graphicx}

\title[PNe as probes for chemical evolution] 
{Planetary nebulae as probes for galactic chemical evolution}

\author[R.D.D. Costa and W.J. Maciel]   
{Roberto D.D. Costa$^1$ \and Walter J. Maciel $^1$}%

\affiliation{$^1$Departamento de Astronomia, IAG, Universidade de S\~ao Paulo,\break
Rua do Mat\~ao 1226, 05508-900, S\~ao Paulo/SP, Brazil \break email: roberto@astro.iag.usp.br; maciel@astro.iag.usp.br

}

\pubyear{2006}
\volume{234}  
\jname{Planetary Nebulae in Our Galaxy and Beyond}
\editors{M.J. Barlow \& R.H. M\'endez, eds.}
\begin{document}

\maketitle

\begin{abstract}
The role of planetary nebulae as probes for the galactic chemical evolution is reviewed. Their abundances
throughout the Galaxy are discussed for key elements, in particular oxygen and other alpha elements.
The abundance distribution derived from planetary nebulae leads to the establishment of radial abundance
gradients in the galactic disk that are important constraints to model the chemical evolution of the Galaxy. 
The radial gradient, well determined for the solar neighborhood, is examined for distinct regions. 
For the galactic anticenter in particular,
the observational data confirm results from galactic evolution models that point to a decreasing in the
gradient slope at large galactocentric distances. The possible time evolution of the radial gradient is also
examined comparing samples of planetary nebulae of different ages, and the results indicate that a flattening
in the gradient occurred, which is confirmed by some galactic evolution models. 
The galactic bulge is another important
region whose modeling can be constrained by observational results obtained from planetary nebulae. Results
derived in the last few years indicate that bulge nebulae have an abundance distribution similar to that of disk
objects, however with a larger dispersion.

\end{abstract}

\firstsection 
\section{Introduction}

Planetary nebulae (PN) constitute an important tool to the study of the chemical evolution of the Galaxy, 
being present in the disk, bulge and halo populations. By providing accurate abundance determinations of 
several chemical elements, these nebulae offer the possibility of studying both the light elements produced 
in low mass stars, such as helium, carbon and nitrogen, and also heavier elements, such as oxygen, 
sulfur and neon, which result from the nucleosynthesis of massive stars. The first group has abundances 
modified by the evolution of the PN progenitor stars, while the second reflects the conditions of the 
interstellar medium at the time the progenitors were formed.

Abundances derived for PN samples from different structures of the Galaxy can and should be used as
constraints to galactic evolution models in order to reproduce their chemical evolution. In the last few years new samples
from the bulge, disk and halo have been made available in the literature. Together they constitute a set of
abundances that can be used to investigate the chemical evolution of the intermediate mass population
in each galactic substructure.

Examining their chemical composition, important constraints for the modeling of the chemical evolution 
of the Galaxy can be derived, such as the abundance distribution in each galactic substructure, the radial 
gradient of abundances found in the disk, as well as its variation along the galactocentric 
radius and with time. 

In this paper we review some of the work done since the last IAU PN Symposium concerning the applicability
of PN as probes for chemical evolution of the Galaxy. For some interesting discussions of PN as a chemical evolution
tool for other galaxies, the reader is referred to the recent volume edited by \cite{stanghellini}.

\section{The galactic bulge}

From the observational point of view, it should be pointed out that although many new objects with 
accurate positions have been found in the bulge region (see for example \cite{jacoby04}), 
the total number of bulge PN with accurate abundances is still small. Furthermore, a correlation between 
the chemical abundances and bulge kinematics is still to be determined, in contrast with the observed 
properties of the galactic disk. Besides, due to the nature and short lifetimes of PN, the presently 
available observational results are strongly biased since they are focused on brighter and younger objects. 

In the past few years, many papers have been published dealing with the kinematics and abundances of the galactic 
bulge. Most of them are concerned with heavy elements produced by supernovae, so that 
light elements such as helium and nitrogen have had a smaller share of attention. In this context, PN constitute 
an important tool in the study of the bulge chemical evolution, providing accurate determinations of the 
abundances of light elements produced by progenitor stars of different masses, as well as heavier elements which 
result from the nucleosynthesis of large mass stars.

\begin{figure}
 \includegraphics[height=2.5in]{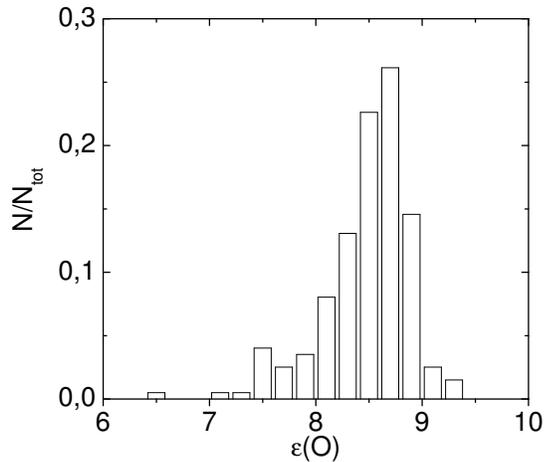}
  \caption{Distribution of oxygen abundances for bulge PN. (Escudero et al. 2004)}
\label{fig:distribution}
\end{figure}

Galactic bulge PN provide accurate abundances of several elements 
that are difficult to study in stars (see for example \cite{esc01}, \cite{esc06}, \cite{exter04}, \cite{gorny04} 
for some recent studies). These data on bulge 
PN can be also used in order to establish observational constraints to the [O/Fe] x [Fe/H] relation for the 
galactic bulge, which is important do derive the bulge evolution, as described by 
\cite{maciel99}. The results derived from different samples are similar. 
They have an abundance distribution similar to the disk with, perhaps, a larger dispersion. However, neither very poor
nor super-metal-rich objects are found. Figure~\ref{fig:distribution} shows the the oxygen abundance distribution 
for bulge PN as given by \cite{esc04}. In this figure and hereafter the notation $\varepsilon$(X)=log(X/H)+12 was adopted.

\begin{figure}
 \includegraphics[height=2.5in]{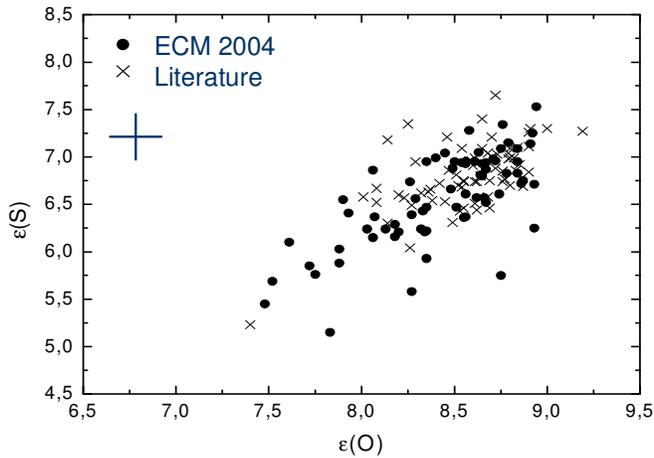}
  \caption{Sulfur abundance vs. oxygen abundance for bulge PN. Dots represent the sample of \cite{esc04}
  and crosses the sample of \cite{exter04}. It can be seen that both samples have similar distribution. (adapted 
from Escudero et al. 2004)}\label{fig:s_o}
\end{figure}

Figure~\ref{fig:s_o} illustrates the typical distance-independent correlations between abundances. 
Dots represent the sample of \cite{esc04} and crosses other data from the literature, mainly from \cite{exter04}.
The average error bar for the abundances is displayed at the upper left corner.
The figure correlates the sulfur and oxygen abundances, which reflect the chemical abundances in 
the interstellar medium at the progenitor formation epoch. It can be seen that both samples have quite similar
distributions. In this kind of diagram, high, intermediate and low mass objects are distributed sequentially, 
with higher mass objects located near the upper right corner of the figure while lower mass objects are at the
lower left corner.

Additional results can be found investigating a special subsample of bulge PN: those with Wolf-Rayet central stars. 
\cite{gorny04} investigate 164 PN toward the bulge, merging data from the literature with new observations, 
finding 18 objects with WR central stars and 24 other central stars with weak emission lines. 
Investigating these objects they found no difference between oxygen abundances of normal PN and those with WR 
central stars, which implies that the oxygen abundances in WR are not significantly affected by nucleosynthesis
or mixing in these objects.

\begin{figure}
 \includegraphics[height=3in, angle=-90]{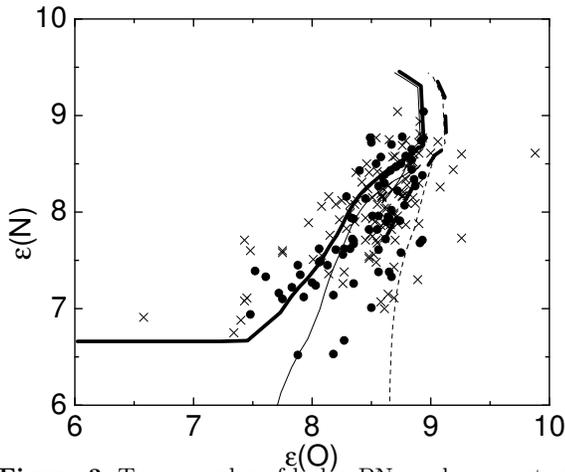}
  \caption{Two samples of bulge PN used as constraints for a bulge chemical evolution model. Continuous and dashed
  lines correspond to different zones in the bulge and continuous, and thick and thin lines correspond to
  different yields for low mass stars. (Escudero et al. 2006)}\label{fig:model}
\end{figure}

As representatives of the intermediate age population of the bulge, PN can be used as constraints for chemical
evolution models of this region. As an example, in figure~\ref{fig:model} two samples of bulge PN are compared to different outputs of a 
multizone, double-infall chemical evolution model of the bulge from \cite{esc06}. Continuous lines represent the
central region of the bulge and dashed lines correspond to the first ring 1.5 kpc away. Thick and thin
lines represent model outputs using different yields for low mass stars.
It can be seen that the intermediate mass population described by the model fits reasonably well the observational
data derived from PN. A small group of objects with
low nitrogen and high oxygen abundances appears in both samples. Its presence can be understood if they are produced
by a second infall episode, when material previously enriched by type-II supernovae ejecta falls onto the disk.

\section{The radial gradient}

The existence of a radial gradient of abundances in spiral galaxies is known for a long time. Studying disks
of spiral galaxies, \cite{searle} pointed out that abundance gradients should not be confined to the inner
regions of galaxies but extend right across galactic disks. As in other spirals, many results report the existence
of a radial gradient in the Milky Way in the solar neighborhood (see \cite{macos03} or \cite{henry} for previous
reviews). These results can also be examined within a more general framework of chemical evolution of galaxies,
as made by \cite{pagel}. 

In the last years several works appeared in the literature, defining the radial gradient from different
objects. Some of the recent works include \cite{daflon} (using OB stars), \cite{deharveng} and \cite{carigi}
(using HII regions), \cite{andrievsky} (using cepheids), \cite{friel} (using open clusters) and \cite{malacos05}
(using PN). In these works the derived slope for the gradient typically varies between -0.03 and -0.07 dex/kpc, as can be seen
in the compilation by \cite{stasinska}, which includes works from different authors, using PN and HII regions to
derive the radial gradient from several elements. Using a homogeneous sample of PN, \cite{henry04} find slopes
between -0.03 and -0.05 dex/kpc for different elements.

For PN in particular, some key problems arise in the derivation of gradients. Their distances are usually derived
from statistical methods, and errors are within a factor of two or even more. There are different distance scales
available in the literature but, despite the fact that the computed slope in principle depend on the adopted
distance for each object, the choice between different statistical distances does not significantly affects the
magnitude of the gradient. Other important points that have to be taken into account when deriving gradients
are the homogeneity of the samples and the relationship between nebular abundances and stellar nucleosynthesis for 
those elements whose abundances are changed during the progenitor evolution.

\subsection{The anticenter}

Most of the results concerning the radial gradient of abundances consider samples of objects with galactocentric
distances between 3 and 10 kpc. Generally there is a considerable under-sampling of objects with distances larger
than 10 kpc, and when the outer part of the disk is considered, results derived from PN indicate a flattening in
the gradient around 10 kpc, as shown by \cite{costa04}. This result is also supported by data derived from cepheids,
however there are contradictory results when HII regions and OB stars are used. Theoretical models for the 
evolution of spiral galaxies, e.g. \cite{chiappini03} or \cite{molla}, also support the existence of a flattening 
in the radial gradient of abundances around 10 kpc.

At present, it is not possible to know what happens to the gradient $beyond$ 10 kpc. Data from PN and cepheids
indicate that there is a flattening indeed, but beyond this point the samples are too small to clearly define either a
slope or a constant value. When examining the chemical evolution models for the galactic disk, many of them predict
a flattening in the abundances; however a different result appears when chemical and dynamical evolution
of the galactic disk are combined. As described by \cite{mishurov}, a chemo-dynamical model indicates a possible 
minimum of abundances around 8-9 kpc from the galactic center, followed not by a flattening but by an inversion in
the gradient slope, with abundances increasing again outward. This effect could be explained by the presence of the 
co-rotation radius of the Milky Way nearly 8.5 kpc from the center. At this position there would be no compression
of the spiral pattern onto the disk, causing a minimum in the star formation and then a minimum in the radial abundance
distribution. Clearly this hypothesis deserves further attention, and new, larger samples of abundances at large galactocentric 
distances are required to better verify this hypothesis.
 
The flattening of the abundance gradient beyond 10 kpc probably reflects a lower star formation rate in the 
outer part of the disk compared to the regions closer to the galactic center. It is apparent from distinct PN data 
samples but less apparent, or
even inexistent, from O,B stars and HII regions, which implies a possible correlation between flattening
and the time variation of the gradient.

\subsection{The time variation of the radial gradient}

As demonstrated by \cite{macouc03}, if a sample of disk PN is divided in groups according to the 
approximate ages of their progenitor stars and the 
radial gradient is derived for each group, younger objects will always tend to have a flatter gradient, no matter 
how the age limits between the groups are defined. Based on a sample of PN in the galactic disk, they concluded that
the O/H gradient flattens out from -0.11 to -0.06 dex/kpc during the last 9 Gyr, or from -0.08 to -0.06 dex/kpc 
in the last 5 Gyr. Later, \cite{malacos05} extended this analysis including S/H data and deriving the [Fe/H] 
gradient based on [Fe/H] $\times$ [O/H] and [Fe/H] $\times$ [S/H] calibrations, confirming the previous results. 
A detailed analysis of the errors involved in the determination of the gradients has been given by \cite{malacos05b}

\begin{figure}
 \includegraphics[height=3in, angle=-90]{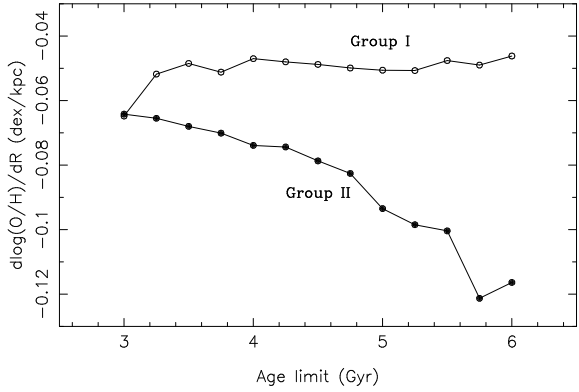}
  \caption{Time variation of the O/H gradient from planetary nebulae. The PN sample was divided in two age groups,
Group I (`younger'), with ages lower than the age limit, and Group II (`older'), with ages higher than the limit. 
The plot shows the O/H gradient (in dex/kpc) of each group as a function of the age limit separating the groups. It can be seen
that the gradients of the younger Group I are always flatter than those of the older Group II. 
(Maciel et al. 2005a)}\label{fig:timevar}
\end{figure}

This approach is illustrated in figure~\ref{fig:timevar}. Group I (empty
circles) is younger, with ages lower than the age limit shown, and Group II (filled circles) is older, with ages
higher than the limit. The age limit that separates the groups varied from 3 to 6 Gyr. For each
age limit, in steps of 0.25 Gyr, the gradient was calculated for both groups. The result indicates that the younger
group always has a flatter gradient.

\begin{figure}
 \includegraphics[height=4in, angle=-90]{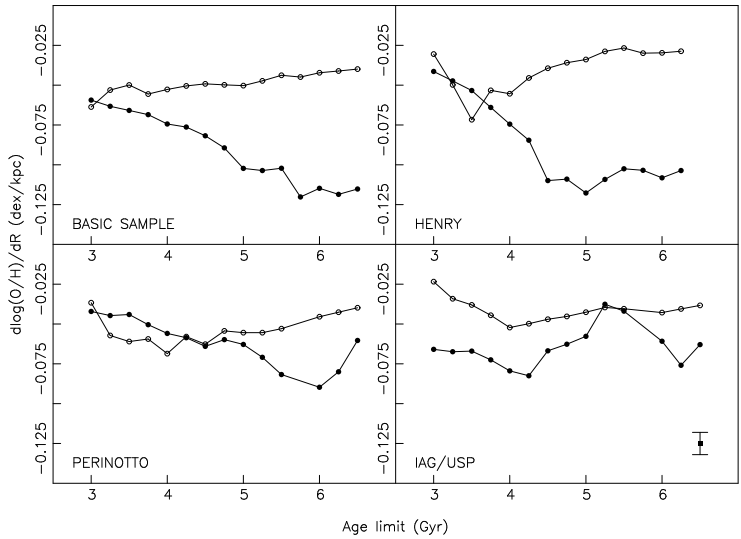}
  \caption{Time variation of the O/H gradient displayed for four PN samples. As in figure 4, 
the samples were divided in two age groups. The age limit between the groups varies from 3 to 6 Gyr. 
(Maciel et al. 2006)}\label{fig:4samples}
\end{figure}

More recently, \cite{malacos06} extended this discussion with data for oxygen, sulfur, argon and neon, and 
comparing four distinct PN samples, including both highly homogeneous samples and compilations. 
Examining them it was possible to verify that the derived
results are essentially the same, in the sense that younger objects show a flatter gradient, no matter the source of the 
abundances. This result is illustrated in figure~\ref{fig:4samples}, where four different samples of PN where used,
showing basically the same result, in spite of some fluctuation that can be attributed mainly to under-sampling. The
average error bar for the oxygen abundances is displayed at the lower right corner. The four samples used to derive
these results are discussed in detail in the paper above mentioned, where similar plots for other elements are also presented. 

\begin{figure}
 \includegraphics[height=4in, angle=-90]{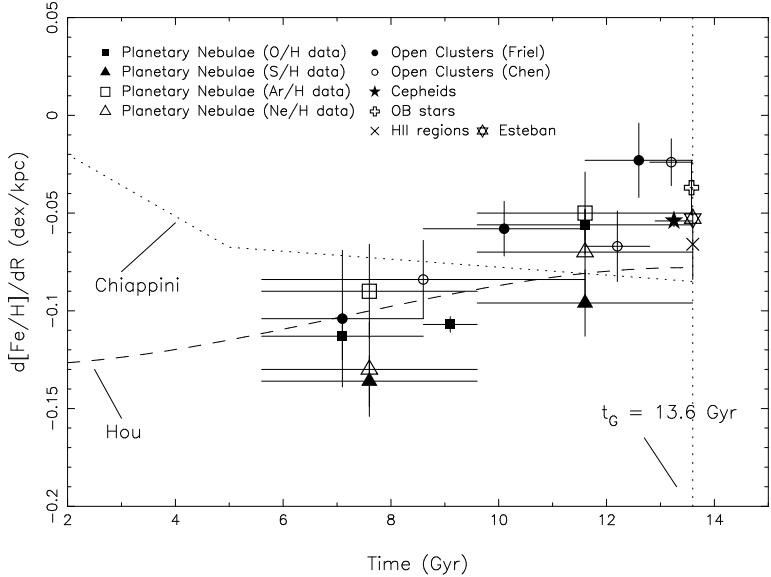}
  \caption{The magnitude of the radial [Fe/H] gradient derived from different objects, reflecting the abundances
of the interstellar medium at distinct epochs. Gradients from PN are combined with data from open clusters, 
cepheids, OB stars and HII regions, and the result indicates that older objects have gradients with larger slopes than those 
derived from HII regions or OB stars. The outputs of two chemical evolution models are also shown,
and the adopted age for the galactic disk is indicated at the lower right corner.
(Maciel et al. 2006)}\label{fig:compare}
\end{figure}

Figure~\ref{fig:compare}, from \cite{malacos06} summarizes the conclusions on the temporal variation of the radial 
gradient. It displays the magnitude of the radial [Fe/H] gradient derived from different objects, reflecting the abundances
of the interstellar medium at distinct epochs. Data for PN are divided in `young' and `old' groups, and combined with
additional data from open clusters, cepheids, OB stars and HII regions. The result indicates that older objects display gradients
with larger slopes than those derived from HII regions or OB stars, which reflect the present abundances of the galactic
disk. In the same figure chemical evolution models for the Galaxy by \cite{chiappini01} and \cite{hou} are also shown
as illustrations. It can be seen that the latter, which predicts a time flattening of the radial gradient, fits 
well the observational data.  

The origin for the time variation of the radial gradient is related to the evolution of the galactic disk. 
It can be understood considering that the flattening in the gradient, initially located near the
outer border of the galactic disk, spreads inward as the galactic chemical evolution proceeds, 
resulting in a flatter gradient. 
Gradients that flatten out in time with a decreasing flattening rate in the last
few Gyr are supported by models proposed by \cite{hou} or \cite{molla}. In particular, different
timescales for star formation and infall rates can account for the detected time variation of the radial gradient.
In fact, models such as those by \cite{hou} or \cite{chiappini01} can predict either flattening or steepening of
the gradient, depending on the star formation and infall timescales.

\section{The halo}

Halo PN belong to the old, metal poor halo population, and therefore can be used to probe the chemical
evolution of its intermediate mass population, in the same way PN of types I, II, III are used to infer the
disk chemical evolution. 

However, the available sample of halo PN is very small. Collecting data up to the end of 2001, \cite{stasinska}
mentions a total of 20 objects. \cite{howard} derived the chemical composition for nine halo PN, finding
subsolar O/H abundances for the whole sample, with $\varepsilon$(O)=7.61 for K648, the most metal poor object of
their sample. They also found that the spread in Ne/O, S/O and Ar/O were consistent with the scatter found
for halo stars, suggesting that accretion of extragalactic material occurred during the halo formation. 

In the last few years some additional objects had their abundances reported. One of them deserves particular attention:
PN G 135.9+55.9 was studied by \cite{richer}, who found extremely low oxygen abundances, 5.8 $<$ $\varepsilon$(O)
$<$ 6.5, with H$\alpha$/H$\beta$ unusually low, and found to be variable between different runs and even among
individual spectra, suggesting the presence of some sort of accretion disk. Later \cite{pequignot} examined more
accurately the atomic physics involved in the chemical diagnosis and concluded that the oxygen abundance is 
around 1/30 solar, which is the lowest oxygen abundance for a planetary nebula. They concluded that this
is an extreme Population II object, reinforcing the idea of an endogenous origin of part of the oxygen. 

\begin{acknowledgments}
This work was partly supported by the Brazilian agencies CNPq and FAPESP.

\end{acknowledgments}

\end{document}